\documentclass[preprint2]{aastex}

\shorttitle{THE CRITICAL ROTATION OF STRANGE STARS}
 \shortauthors{Zheng et al.}

\begin{document}


\title{ THE CRITICAL ROTATION OF STRANGE STARS AND RAPIDLY ROTATING PULSARS}


\author{Zheng Xiaoping  and Yang Shuhua}
\affil{Department of Physics, Huazhong Normal University, Wuhan
430079}
 \email{zhxp@phy.ccnu.edu.cn}
\author{Li Jiarong}
\affil{The Institute of Particle Physics, Huazhong Normal
University, Wuhan 430079}


\begin{abstract}
We utilize the bulk viscosity of interacting strange quark matter
to reevaluate the damping time scale. The presence of medium
effect of bulk viscosity leads to a stronger damping of r-modes,
which can be over an order of magnitude for realistic parameters.
We find that the r-mode instability window is narrowed due to the
medium effect, and hence when a pulsar reaches the instability
window it will only slow down by gravitational wave emission to a
period of 1.78msec instead of 2.5msec given by early estimate. As
a theoretical upper rotation limit of pulsars, the period of
1.78msec is very close to the two most rapidly spinning pulsars
known, with periods of about 1.6msec.
\end{abstract}




 \section {INTRODUCTION}
At high density, normal hadron matter has been predicted a
deconfined transition, where quark-gluon plasma is formed. This
could have important consequences for compact stars, with central
densities several times the nuclear saturation density. Ever since
strange quark matter(SQM), a conglomerate of up, down and strange
quarks, has been suggested as a possible absolutely stable or
metastable phase of nuclear matter\citep{wit84}, it has been
speculated that strange stars might exist in the
universe\citep{alc86, hae86, col92}.

The studies that quark interactions within lowest order
perturbative QCD were considered in the MIT bag model also
predicted that sufficiently heavy strangelets might be absolutely
stable\citep{fah84}. If the SQM hypothesis is correct then some
(perhaps all) pulsars may be strange stars\citep{alc86, hae86,
col92}. Thus it would have obvious implications for our
understanding of pulsars if strange stars could exist. It is an
important subject that probing possible observational evidences of
the existence of strange stars by astrophysical investigations.

Since self-bound of SQM due to strong interaction is very
important, the mass-radius relation of the assumed strange stars
has been uncovered to be different from that of neutron stars.
However, for the canonical mass of $1.4M_{\odot}$, gravity
dominates the strong interaction which leads to strange stars and
neutrons being similar in size\citep{alc86}. Therefore, the
attempt to distinguish strange stars from neutron stars for given
observed masses and radii of pulsars seems to have encountered
with great difficulties. Strange stars could indicate the
distinguishable signal from neutron stars by their cooling
properties, but the difference in the cooling behavior of strange
stars and neutron stars will disappears if the direct Urca process
is operating in the core of an neutron star, which is done either
by the proton fraction rising above $~10\%$ \citep{lat94} or by
hyperons being present in the core of a neutron star\citep{pag00}.
This would make it difficult to identify a strange star from
cooling data.

Ever since Andersson realized that the r-modes are unstable at all
rates of rotation in perfect fluid stars\citep{and98}, a series of
papers have investigated the many implications for gravitational
radiation detection and the evolution of pulsar\citep{fri98,
lin98, owe98, koj98, mad98, and99a, and99b, ho00, rez00,
mad00,and02}. In recent years, several crucial issues regarding
the astrophysical relevance of the r-mode instability have been
investigated. Key results concern the interaction between
oscillations in core fluid and the crust\citep{bil00, and00,
lin00a}, the role of the magnetic field\citep{spr99, rezz00,
men01} superfluidity\citep{lin00b,and01} and the effect of exotic
particles that are thought to exist in the deep neutron star
core\citep{jon01, lin01}.

Meanwhile, \citet{mad98} has pointed out that the r-mode
instability may provide the means to distinguish strange stars
from neutron stars. The main reason for this is that the viscosity
coefficients are rather different in these two cases. While the
shear viscosity of a strange star is comparable to that of a
neutron star, the bulk viscosity would be many orders of magnitude
stronger than its neutron star counterpart. This has interesting
effects on the r-mode instability. Based on the characteristic
r-mode instability window of strange stars, which is related to
the gravitational wave emission and the viscosity, a few
mechanisms have been discussed to explain the clustering of
spin-frequencies of LMXBs\citep{mad00, and02}.

However, the bulk viscosity coefficient in Madsen's studies takes
the one for the case of non-interacting quark gas in MIT
bag\citep{mad92}. In fact, if the interaction among quarks is
considered, the bulk viscosity increases, which is calculated by
\citet{zhe02}. As we will show in the following, the r-mode
instability window for strange stars will be significantly
modified, when the improved bulk viscosity is adopted.

As demonstrated blow, we find the r-mode instability window is
evidently narrowed due to the inclusion of medium
effect(interactions among constituent particles). The medium
effect makes the lowest critical spin frequency  rise to 558Hz
from about 300-400Hz(for the case of free quark gas). The limiting
period of 1.78msec corresponding 558Hz is closer to the two most
rapidly spinning pulsars known (with the periods of 1.56msec and
1.61msec) than the periods of 2.5msec and 3msec given by Madsen.

In addition, if it could be established that rapidly spinning
pulsar are strange stars, the pulsar data would put certain
constraints on the model parameter, namely the current mass of
strange quark mass $m_{s}$. As we can see in Madsen's
article\citep{mad00}, in the normal and especially in the 2-flavor
color superconducting phase (2SC) case, the theoretical results is
not consistent with the pulsar data when $m_{s}=100$MeV. However,
when we taking the medium effects into account, the constraints to
$m_{s}$ from pulsar data appears to be deeply relaxed, and we will
show that $m_{s}$ can be taken as 100MeV in this article.

This paper is organized as the following. In section 2, we recall
the bulk viscosity of interacting quark matter and compute the
damping time scale on r-modes. In section 3, we give the improved
critical curve(spin frequency) in spin
frequency($\nu$)-temperature($T$) plane. In section 4, we  give a
summary.

\section{BULK VISCOSITY AND DISSIPATION ON r-MODES}
    Ever since  \citet{wan84} pointed out that strange quark matter
is characterized by a huge bulk viscosity relative to nuclear
matter, some investigations have tried to calculate the relevant
viscosity coefficient of SQM\citep{saw89, mad92, goy94}. In a MIT
bag model, it is thought that the exact solution had been obtained
by Madsen\citep{mad92}. However, he ignored the coupling among
quarks in the bag, which had  been considered in the study of the
equation of state of SQM\citep{sch97}. Soon before, we found the
coupling's effect on the bulk viscosity leads to an increase of
the viscosity over an order of magnitude although the medium
modifications of the equation of state of strange quark matter was
proven to be small\citep{zhe02}. To very good approximation, the
relevant bulk viscosity coefficient still takes the form
formulated by \citet{mad92}:
\begin{equation}
  \zeta =\frac{\alpha T^{2}}{\omega^{2}+\beta T^{4}}
\end{equation}
but here $\alpha$ and $\beta$ were given by \citep{zhe02}, which
are extremely  differing from Madsen's and strongly depending on
the coupling constant of strong interactions among quarks. We
expressed them as
\begin{equation}
\alpha =9.39\times 10^{22}\mu _{d}^{5}\left( \frac{k_{{\rm
F}d}^{2}}{ C_{d}}-\frac{k_{{\rm F}s}^{2}}{C_{s}}\right) ^{2}{\rm
(gcm}^{-1}{\rm s}^{-1}),
\end{equation}
\begin{equation}\beta =7.11\times 14^{-4}\left[ \frac{\mu _{d}^{5}}{2}\left( \frac{1}{k_{
{\rm F}d}C_{d}}-\frac{1}{k_{{\rm F}s}C_{s}}\right) \right]
^{2}{\rm (s}^{-2}),
\end{equation}
 where
$k_{Fi}=(\mu_i^2-m^{*2}_i)^{1/2},C_{i}=\mu
_{i}-m_{i}^{*}\frac{\partial m^{*}_i}{\partial \mu _{i}}$ and
$m_i^{*}$ was given in the references\citep{kli82, wel82, pis89,
bla93, vij95}.

The bulk viscosity is expected to be the dominated internal fluid
mechanism in hot compact stars. The timescales for the bulk
viscosity damping of r-modes need to be estimated. Since the
coupling of the r-modes to bulk viscosity vanishes in the
lowest-order expression, the derived timescales must be based on
the fully self-consistent second-order calculation of this
coupling. The earlier estimates can't gain this end\citep{lin98,
and99a, kok99}. But it has been completed by \citet{lin99}. We
substitute our viscosity for the given Low-T limit($T<10^{9}K$)
into the formula(6.2)given by \citet{lin99}, and find the only
difference comparing to the early calculation is the change of the
viscosity coefficient. Thus, we can immediately obtain the
timescale by a simple comparison instead of the repeat of the
complicated calculations
\begin{equation}
  \tau_{B}=\bar{\tau}_{B}{\rm s}(\pi
  G\bar{\rho}/\Omega^{2})T_{9}^{-2},
\end{equation}
with
\begin{equation}
\bar{\tau}_{B}=2.83 \times 10^{3} \alpha^{-1} \bar{\rho}
m_{s100}^{4}.
\end{equation}
 where $T_9$ and $m_{100}$ denote temperature in units of $10^9$ K and the  current mass of strange quark
 in units of 100 MeV  ,
 and $\bar{\rho}$ is the mean density of the star.

 Evidently, $\bar{\tau}_B$
 is  determined with the chemical potential $\mu_d$, which can be
 obtained by solving the equations related to chemical
 equilibrium, electric charge neutrality and conservation of
 baryon number for given $\bar\rho$, namely
 \begin{equation}\label{a}
 \mu_{s}=\mu_{d}, \mu_{u}=\mu-\mu_{e}
 \end{equation}
\begin{equation}\label{b}
\frac{2}{3}n_{u}-\frac{1}{3}(n_{d}+n_{s})-n_{e}=0
\end{equation}
 \begin{equation}\label{c}
 n=\frac{1}{3}(n_{u}+n_{d}+n_{s})
 \end{equation}
 \begin{equation}\bar{\rho}=(\frac{E}{A})n
 \end{equation}
 Here, $n$ denotes the baryon number density, and
 $n_{i}=\frac{1}{6\pi^{2}}k_{Fi}^{2}$ is the particle number
 density. $\frac{E}{A}$ is the energy per baryon and we approximately
 take it as the mass of a neutron in our calculation.

The coupling $g$ will extremely influences the solutions of the
above equations. Following \citet{sch97}, $g$ is taken as a free
parameter ranging from 0 to 5 and the equations(6) (7) (8) are
solved numerically. Figure 1 shows $\bar{\tau}_B$ as a function of
$\bar\rho$ for different $g$ and $m_s=200$MeV. The time scales
depend on $g$ remarkably while they are nearly independent of
$\bar\rho$.  The time scale for $g=5$ arrives at tens of times
shorter than the case of $g=0$. For a strange star with mass
  $M=1.4M_\odot$ and radius $R=10$km, we find $\bar{\tau}_B$
  ranges from
  4.24$\times 10^{-2}$s$(g=0)$ to 1.53$\times 10^{-3}$s$(g=5)$. This
  is of great interest because this will significantly increases the critical
  rotation angular velocity for the onset of r-mode instability.
\section{CRITICAL SPIN FREQUENCY FOR THE ONSET OF THE r-MODE
INSTABILITY}
   The r-mode unstable (or stable) regime of the relativistic stars,
neutron stars as well as strange stars,  depends on the
competition between the gravitational  radiation and various
dissipation mechanisms. To plot the instability window of r-mode
or gain the critical rotation frequency for a given stellar model
as a function of temperature, we need to acquire the
characteristic timescales, damping and growing timescales of
r-mode instability. Due to the emission of gravitational waves,
the r-mode grows on a timescale
\begin{equation}
\tau_{\rm G}=\bar{\tau}_{\rm G}({\rm\pi}G\bar\rho/\Omega^2)^3
\end{equation}
where $\bar{\tau}_{\rm G}$ is $-3.26(1.57)$s for $n=1(0)$
polytropic EOS, which was respectively studied by \citet{lin99}
and \citet{kok99}. For the viscous damping timescales, we shall
also consider the shear viscosity besides bulk viscosity discussed
in the last section. In strange stars,  the time scale for the
shear viscous damping is given
\begin{equation}
\tau_S=\bar{\tau}_S(\alpha_S/0.1)^{5/3}T_9^{5/3}
\end{equation}
Here, $\bar{\tau}_S$ is $5.37(2.40)\times 10^8$s corresponding to
$n=1(0)$, $\alpha_S$ is the strong coupling and we take
$\alpha_S=0.1$ in following calculation\citep{mad00}.

We can now evaluate the critical spin frequency as a function of
temperature  from the equation
\begin{equation}
{1\over\tau_{G}}+{1\over\tau_{S}}+{1\over\tau_{B}}=0
\end{equation}

  Figure 2 shows the regions of r-mode (in)stability in spin
  frequency-temperature($\nu-T$)plane for a strange star with mass
  $M=1.4M_\odot$ and radius $R=10$km. The shading between the two
  curves displays the effect of the medium modification  of quark masses on
  critical rotation frequencies. The medium effect  narrows the r-mode instability window.
  The dotted curve  corresponds to  $g=0$,
  reduced to the Madsen's result\citep{mad00}, and $g=5$ is assumed in the
  upper  contour(the solid curve).  The upper contour has a lowest limiting
  frequency denoted by $C$
  as 558Hz(the corresponding period is 1.78msec), which is
  more close to the two most rapidly spinning
  pulsars known, with frequencies of 642Hz and 622Hz( the periods
  are 1.56 and 1.61msec), than the period of 2.5msec. This implies a strange star would slow down
  by gravitational wave emission when it reaches the instability window and spin around in
  1.78msec instead of the 2.5$\sim$3msec expected by Madsen.
  Figure
  3 shows
   the results for 2SC
   stars. Similar to figure 1, the medium effect also increases the critical
   rotation frequency of 2SC stars, but it should be stressed that the medium
   effect leaves the most rapidly stars away from the instability window.

    Figure 4 depicts the instability windows where the current mass of
strange quark takes 100 MeV. If the rapidly rotation pulsars could
be regarded as strange stars, the medium effect would relax the
stringent constraint  on the choice of QCD parameters in contrast
to the non-interacting medium case, for example, the 2SC stars for
smaller current mass $m_s$ can safely exist due to the medium
effect(see the upper dashed curve in figure 4).
\section{CONCLUSIONS AND DISCUSSIONS }
       We apply the bulk viscous coefficient including medium
effect to reevaluate and discuss the viscous damping time. The
time greatly depends on the strong coupling $g$. We find the time
scale $\bar{\tau}_B$, rather than non-interacting perfect fluid
where $\bar{\tau}_B$ is constant(The fact is that $\bar{\tau}_B$
versus $\bar{\rho}$ slowly increases for non-vanishing $m_s$ as
shown in figure 1), weakly decreases with increasing mass density
of stars for the given larger $g$. Therefore, the medium effect
due to strong interactions among quarks adds the viscous
dissipation of instability modes. For a star with $M=1.4M_\odot$
and radius $R=10$km, we numerically calculate the critical spin
frequency as a function of temperature. The medium effect lead  to
a significant lift of the critical frequencies. We find that the
largest critical period is reduced to 1.78msec from 2.5msec,
appearing to be more close to  the two most rapidly spinning
pulsars known.

We also study the 2SC stars and give our improved scenario. The
current mass of strange quark in the model have a broader range
taking the medium effects into account.

Finally, it should be mentioned that we here only consider weak
coupling of strong interactions among quarks in MIT bag. The long
range unperturbation interactions are contained in the usual bag
constant. Although the MIT bag constant is able to contribute to
EOS of SQM, the long range effects on dynamical quantities, such
as viscous coefficient etc., seem to be ignored due to the hiding
of the microscopic processes into the bag constant. If the
processes can be taken into account, we may conjecture that the
limiting spin of strange stars would shift further  upward. This
is our future work.

\section*{Acknowledgments}
We would like to thank the partial support by National Natural
Science Foundation of China under Grant No. 10175026 and Natural
Science Foundation of CCNU.

\clearpage

\begin{figure}
\plotone{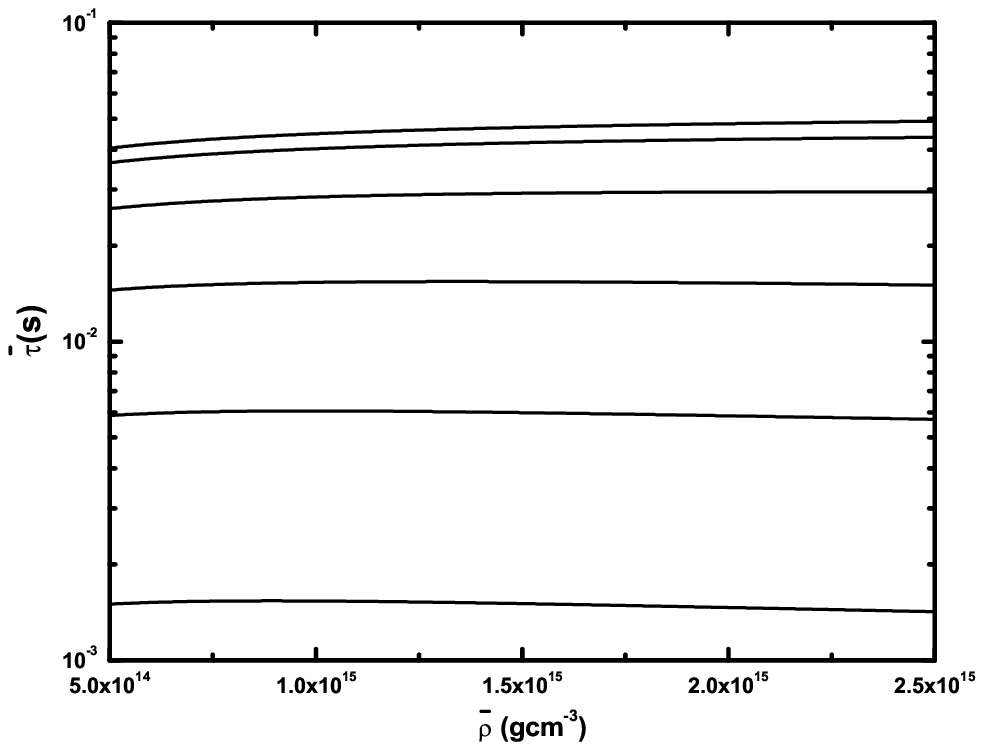} \caption{ The time scales $\bar\tau_B$ as
functions of mass density for different couplings $g$, which are
0, 1, 2, 3, 4, 5 respectively from top to bottom  \label{fig1}}
\end{figure}
\clearpage

\begin{figure}
\plotone{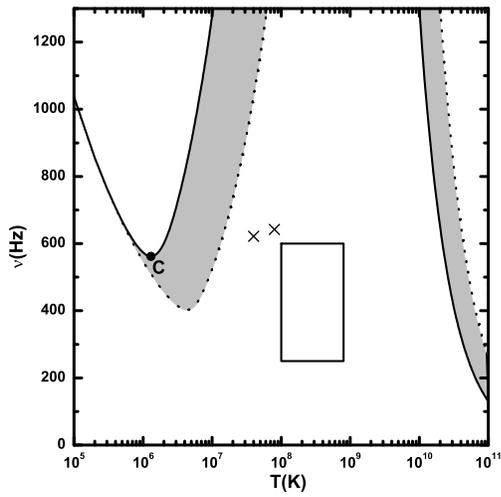} \caption{Critical spin frequencies for strange
stars as functions of temperature. The  lower contour(dotted)
curves stand for free quark gas. The shadow displays the medium
effect. \label{fig2}}
\end{figure}
\clearpage

\begin{figure}
\plotone{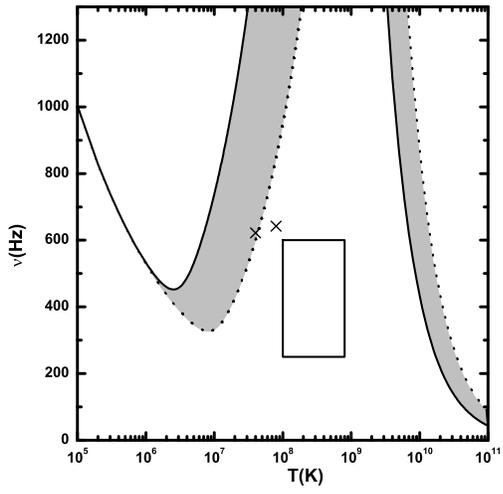} \caption{As Fig 2, but assuming a 2SC state.
\label{fig3}}
\end{figure}
\clearpage

\begin{figure}
\plotone{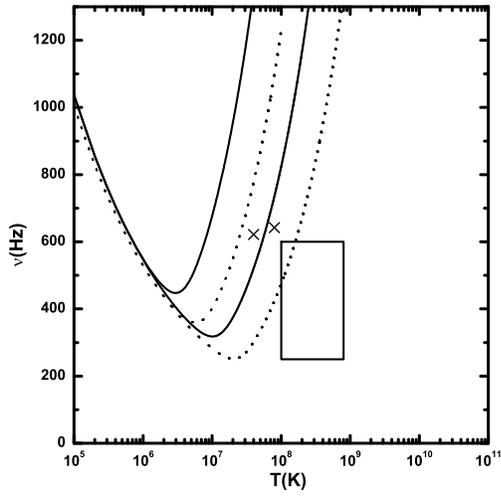}\caption{The r-mode instability windows for the
case of $m_s$=100MeV. The solid curves denote the strange stars
with normal phase and the dashed curves represent the case with
2SC phase. The lower solid and dashed curves correspond to the
case of non-interacting quark matter.\label{fig4}}
\end{figure}
\end{document}